\documentclass[preprint,12pt]{elsarticle}

\usepackage{amssymb}
\usepackage{amsmath}
\usepackage{subcaption}\usepackage{slashed}\usepackage[utf8]{inputenc}  
\usepackage[T1]{fontenc} 
\usepackage{hyperref}
\usepackage{xcolor}

\journal{Nuclear Physics B}

\begin{document}

\begin{frontmatter}



\title{Background Field Effects on Quasi-Real Photon Emission and Lepton-Pair Production at EIC and EicC}

\author[1]{Cong Li}

\affiliation[1]{%
  organization={School of Information Engineering, Zhejiang Ocean University},
  addressline={}, 
  city={Zhoushan},
  postcode={},
  state={Zhejiang},
  country={China}
}

\begin{abstract}
We study how background electromagnetic fields modify quasi-real photon emission at the EIC and EicC through an effective coupling correction, thereby altering the photon flux spectrum. The resulting change in lepton-pair production via photon–photon fusion—where one photon arises from the electron and the other from the nuclear Coulomb field—serves as a clean QED probe of such background effects. In our formulation, the correction originates from a non-perturbative modification of the photon propagator induced by the background field. Because this process shares the same initial-state photon dynamics as photon–gluon fusion, any background-induced alteration of the photon propagator directly impacts the extraction of small-$x$ gluon distributions. Numerical estimates at realistic collider energies indicate that these corrections should be non-negligible in certain kinematic regions.
\end{abstract}


\end{frontmatter}



\section{Introduction}
\label{s.intro}
Precisely determining the gluon distribution at small-x is a central objective of high-energy nuclear physics and a key scientific motivation for the Electron-Ion Collider~\cite{Eicwhitebook,MUELLER1999285,GRIBOV19831,McLerran_1994}. In particular, the process in which an electron emits a photon that subsequently probes the gluon content of the target nucleus via $\gamma^* + g \to q\bar{q}$ provides a clean and powerful channel for accessing the unintegrated gluon distribution function (UGDF)~\cite{Dominguez_2011, Kowalski_2003,Balitsky_1996,Kovchegov_1999,PhysRevD.108.034025,PhysRevD.104.074006,Boroun2024}. In this picture, the electron serves as an efficient source of electromagnetic radiation, while the gluon content is extracted from the final-state correlations of the produced quark-antiquark pair~\cite{BUDNEV1975181,PhysRevD.42.3690,ALTARELLI1977298}. However, the theoretical reliability of such extractions hinges on accurately modeling initial-state QED dynamics, particularly virtual photon emission in nuclear electromagnetic fields.

In this work, we investigate the impact of strong classical background electromagnetic fields—such as those present in  highly charged nuclear environments—on the virtual photon emission from the incident electron. This work is motivated by the fact that even purely QED processes can indirectly affect gluon distribution extractions, due to their shared dependence on initial-state photon radiation~\cite{Collins:1989gx, YENNIE1961379}. Specifically, we consider the two-photon process $e + A \to e + \ell^+\ell^- + A$, in which a quasi-real photon ($q^2 \approx0$) emitted by the electron annihilates with a quasi-real photon from the Coulomb field of the nucleus to produce a lepton pair~\cite{BALTZ_2008,PhysRevD.4.1532}. While this channel does not involve QCD interactions directly, it shares the same photon emission kinematics with the $\gamma^* + g \to q\bar{q}$ process \cite{MUELLER1999285}. Consequently, any modification to the photon propagator—whether from vacuum polarization or background field effects—can impact both channels in parallel. This renders the lepton-pair production process a sensitive probe of initial-state QED corrections under realistic collider conditions \cite{Klein_2017}.

Building on the formalism developed in Ref.~\cite{Li3}, which treats the photon propagator in the presence of a background field \cite{Dittrich:2000zu}, we extend the analysis to the kinematic regimes relevant to the Electron Ion Collider (EIC) in US \cite{ABDULKHALEK2022122447,EICwhitepaper} and the Electron Ion Collider in China (EicC) \cite{EICCwhitepaper}. The scientific case for the U.S.-based Electron-Ion Collider has been strongly endorsed at the national level, recognizing its unique capability to address fundamental questions in QCD and nuclear structure~\cite{NAP25171}. In our approach, the effect of the external field is incorporated nonperturbatively by modifying the photon propagator, a method chosen to accurately capture phenomena in strong background fields, leading to an effective electromagnetic coupling constant, $\alpha_{\rm eff}$, that reflects the influence of the background. This effective coupling is then used to systematically adjust the photon flux spectrum that enters cross-section calculations. Physically, background fields reshape the photon emission phase space and thus modify the differential cross section for lepton-pair production.

To make quantitative predictions, we apply this modified coupling to compute the cross section for the $\gamma^*\gamma^* \to \ell^+\ell^-$ process, including realistic electron kinematics and nuclear form factors \cite{Klein_2017,DEVRIES1987495}. Furthermore, we incorporate Sudakov resummation of soft QED radiation \cite{YENNIE1961379,Collins:1989gx}, which introduces logarithmic corrections that become significant in the semi-hard transverse momentum region relevant for the electron ion collision, where large logarithmic enhancements from soft photon emission need to be resummed for theoretical consistency. Such effects are essential for maintaining theoretical consistency in regions where fixed-order calculations fail to capture the correct asymptotic behavior. Our numerical results, presented for both EIC and EicC, reveal that the background-field-induced coupling modification can lead to measurable deviations in the lepton-pair production cross section. If neglected, these deviations could bias UGDF extractions from photon–gluon fusion, as both processes assume the same initial-state photon flux.

The remainder of this paper is organized as follows. In Section ~\ref{2}, we present the theoretical framework for how background electromagnetic fields modify the photon propagator and introduce an effective coupling constant to account for these corrections. We then apply this formalism to recalculate the cross section for lepton-pair production via two-photon fusion, incorporating Sudakov resummation, and realistic nuclear form factors. Numerical results are also provided for collider setups at both EIC and EicC. In Section ~\ref{3}, we summarize our main findings and discuss their implications for future precision studies of QED and gluon distribution measurements at next-generation electron–ion colliders.

\section{Correction of coupling constant by background field}
\label{2}

In electron-ion collisions, we follow an analogous approach to that used in gluon distribution function extraction to instead determine the photon distribution function. In our method, the incoming electron emits a quasi-real virtual photon. To reduce complexity and because in this limit the background photon field can be accurately described using the equivalent photon approximation, the investigation is restricted to the regime where $q^2 \approx 0$. This photon then annihilates with another quasi-real virtual photon from the peripheral electromagnetic field of a lead nucleus, producing a lepton pair.
\begin{equation}
    \gamma_1(k_1) + \gamma_2(k_2) \to l^+(p_1) + l^-(p_2) \label{eq:annihilation}
\end{equation}
The cross section of this process is sensitive to the photon distribution in the nuclear periphery. Analogously, the gluon distribution function is extracted by studying photon-gluon fusion processes that produce quark-antiquark pairs.

In extracting the photon distribution function, any neglected corrections can hinder the accuracy of the result. We pay special attention to processes in which the initial-state electron radiates a virtual photon within a dense background photon field. We find that such a background modifies the electron's photon emission, altering the cross section of the process and, consequently, the extracted photon distribution function. In vacuum, the photon propagator takes the form $ \langle 0 | A_\mu(x) A_\nu(y) | 0 \rangle.$ However, in the presence of a background field $ |B\rangle $, the photon propagator is modified, as shown below.
\begin{equation}
    \begin{aligned}
\langle B| A_\mu(x) A_\nu(y) |B \rangle
&= \mathrm{Tr}\!\left[\rho\, A_\mu(x) A_\nu(y)\right]  \\[4pt]
&= \sum_\lambda \int \frac{d^3k}{(2\pi)^3 2k_0}\, n(k)\,
\mathrm{Tr}\!\left[ |k\rangle \langle k|\, A_\mu(x) A_\nu(y) \right]  \\[4pt]
&= \sum_\lambda \int \frac{d^3k}{(2\pi)^3 2k_0}\, n(k)\,
\langle k| A_\mu(x) A_\nu(y) |k\rangle  \\[4pt]
&= \int \frac{d^3k}{(2\pi)^3 2k_0}\, n(k)\, e^{-ik\cdot(x-y)}
\sum_\lambda \epsilon_\mu^{(\lambda)*} \epsilon_\nu^{(\lambda)}  \\[4pt]
&= \int \frac{d^3k}{(2\pi)^3 2k_0}\, n(k)\, e^{-ik\cdot(x-y)}\, (-g_{\mu\nu}) \\[4pt]
&= \int \frac{d^4k}{(2\pi)^4}\, e^{-ik\cdot(x-y)}\,
\frac{-i g_{\mu\nu}}{k^2+i\varepsilon}\, n(k).
\end{aligned}
\end{equation}
where, $|B\rangle$ denotes the background photon field, which in the present case corresponds to the peripheral electromagnetic field of a relativistic heavy ion. The background density matrix is defined as
\begin{align}
\rho = |B\rangle \langle B|
= \sum_\lambda \int \frac{d^3k}{(2\pi)^3 2k_0}\,
n(k)\, |k\rangle \langle k|,
\end{align}
where $n(k)$ represents the average occupation number of photons with momentum $k$, i.e., the actual number of photons populating the mode $k$, which characterizes the background photon distribution in momentum space, hereafter simply referred to as the photon distribution of background field.

It is evident that the photon propagator in the background field differs from its vacuum counterpart: the background field effectively modifies the propagator by a convolution with the occupation number function $n(k)$. This indicates that photon emission is influenced by the absolute number of background photons, an effect analogous to stimulated radiation in quantum electrodynamics. Unlike in vacuum, the background-modified photon propagator includes a convolution with the photon distribution function $ n(k)$. This function can be absorbed into the electromagnetic coupling constant to define an effective coupling,
\begin{equation}
    \alpha_{\text{eff}}(k) = \alpha_e [n(k)+1] \label{eq:alpha-eff}
\end{equation}
where $ \alpha_e = 1/137 $ is the QED coupling constant. The `1' term in this expression corresponds to the conventional interaction mediated by virtual particles in the vacuum (i.e., vacuum fluctuations). When the background field is removed, the $n(k)$ vanishes, and the interaction strength reverts to the standard $\alpha_e$. However, in high-density backgrounds, such as the heavy ion peripheral electromagnetic field, the modes are highly populated ($n(k) \gg 1$). In this high-density regime, the vacuum term is negligible, and an effective coupling, $\alpha_{\text{eff}} \approx \alpha_e n(k)$, serves as an excellent approximation. This indicates that the interaction is significantly enhanced by the stimulated processes within the background field. 

\begin{figure}[ht]
	\centering 
	\includegraphics[width=0.8\textwidth, angle=0]{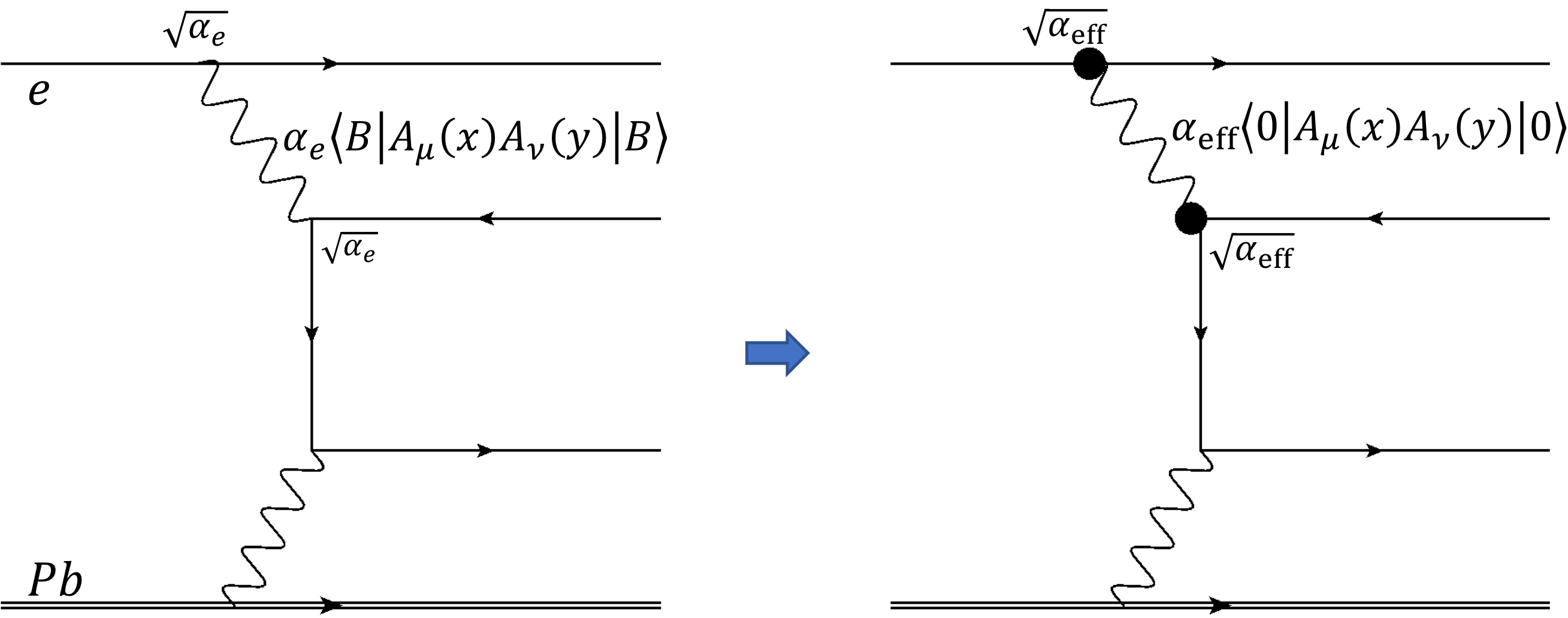}	
	\caption{The effect of the background field on the photon propagator can be equivalently interpreted as a modification to the coupling constant.} 
	\label{11}%
\end{figure}

As shown in Fig.~\ref{11}, the effective coupling $ \alpha_{\text{eff}} $ describes the strength of electromagnetic interactions in the presence of the background field. Here, it’s important to clearly distinguish between the photon propagator $\left\langle B \middle| A_\mu(x)\, A_\nu(y) \middle| B \right\rangle $ and the electron propagator $\left\langle B \middle| \psi(x)\, \bar{\psi}(y) \middle| B \right\rangle=\left\langle x \middle| \frac{i}{\slashed{\hat{p}}-e\slashed{\hat{A}}-m+i\epsilon} \middle| y \right\rangle$. The latter incorporates the familiar Coulomb correction, which we will explore in future work.

Using this effective coupling, we recalculate the cross section for two-photon annihilation into a lepton pair. The essential modification lies in replacing the electromagnetic coupling at the electron vertex with the effective one, as the electron is the source of the emitted quasi-real photon whose propagation is influenced by the background field. The resulting differential cross section is given by~\cite{Li1,Li2,PhysRevD.87.054010,Pisano_2013},
\begin{align}
    \frac{d\sigma}{d^2 p_{1\perp} d^2 p_{2\perp} dy_1 dy_2} 
    &= \frac{2 \alpha_e }{W^4} \frac{W^2 - 2P_\perp^2}{P_\perp^2} \nonumber\\&\times
    \int d^2 k_{1\perp} d^2 k_{2\perp} 
    \delta^2(q_\perp - k_{1\perp} - k_{2\perp}) 
    \nonumber\\&\times\alpha_{\text{eff}}\
    x_1 f_1^{\gamma/e}(x_1, k_{1\perp}^2) x_2 f_1^{\gamma/A}(x_2, k_{2\perp}^2) \label{eq:cross-section}
\end{align}
where $ W $ is the invariant mass of the final-state lepton pair and the total transverse momentum is defined as $ q_\perp = k_{1\perp} + k_{2\perp} = p_{1\perp} + p_{2\perp} \approx 0 $, implying $ p_{1\perp} \approx -p_{2\perp} \approx P_\perp $. In the correlation limit, where two well-separated scales exist, large logarithmic terms emerge from unobserved soft photon radiation in higher-order QED corrections \cite{Klein_2019}. It is conventional to express the cross section in impact parameter space to facilitate the resummation of these logarithms,
\begin{align}
    \frac{d\sigma}{d^2 p_{1\perp} d^2 p_{2\perp} dy_1 dy_2} 
    &= \frac{2 \alpha_e }{W^4} \frac{W^2 - 2P_\perp^2}{P_\perp^2}
    \int \frac{d^2 r_\perp}{(2\pi)^2} e^{i r_\perp \cdot q_\perp} 
    e^{-\frac{\alpha_e}{2\pi} \ln^2 \frac{W^2}{\mu_r^2}} \notag \\
    & \times \int d^2 k_{1\perp} d^2 k_{2\perp}
    e^{-i (k_{1\perp} + k_{2\perp}) \cdot r_\perp} \nonumber\\&\times\alpha_{\text{eff}}\
    x_1 f_1^{\gamma/e}(x_1, k_{1\perp}^2) x_2 f_1^{\gamma/A}(x_2, k_{2\perp}^2) \label{eq:resummed}
\end{align}
where $ \mu_r = 2e^{-\gamma_E}/|r_\perp| $. The electron mass is neglected, which is a valid approximation at EIC and EicC energies. The rapidities of the final-state leptons are denoted by $ y_1 $ and $ y_2 $. By energy-momentum conservation, the momentum fractions of the incoming photons are fixed $k_1^+ = x_1 \bar{P} = |p_{1\perp}| e^{y_1} + |p_{2\perp}| e^{y_2}$ and $k_2^- = x_2 P = |p_{1\perp}| e^{-y_1} + |p_{2\perp}| e^{-y_2}.$

The unpolarized photon distributions in the electron and nucleus are denoted by $ x_1 f_1^{\gamma/e}(x_1, k_{1\perp}^2) $ and $ x_2 f_1^{\gamma/A}(x_2, k_{2\perp}^2) $, respectively. For a boosted nucleus, the photon distribution can be obtained via the Equivalent Photon Approximation (EPA)~\cite{Li1,Li2,eq,eq2},
\begin{equation}
    x f_1^{\gamma/A}(x, k_\perp^2) = \frac{Z^2 \alpha_e}{\pi^2} \frac{k_\perp^2}{(k_\perp^2 + x^2 M_p^2)^2} 
    \left[ F(k_\perp^2 + x^2 M_p^2) \right]^2 \label{eq:photon-nucleus}
\end{equation}
where $ Z $ is the nuclear charge number, $ M_p $ is the proton mass, and $ F $ is the nuclear form factor in momentum space, parameterized from STARlight Monte Carlo~\cite{Klein_2017},
\begin{equation}
    F(|k|) = \frac{4\pi \rho^0}{|k|^3 A} \left[\sin(|k| R_A) - |k| R_A \cos(|k| R_A) \right] 
    \frac{1}{1 + a^2 |k|^2} \label{eq:form-factor}
\end{equation}
with $ R_A = 1.1 A^{1/3} $ fm, $ a = 0.7 $ fm, and $ \rho^0 $ a normalization constant. This form is numerically close to the Woods-Saxon distribution and useful in numerical evaluations~\cite{DEVRIES1987495}. The photon distribution in the boosted electron is given by~\cite{cc3,Vysotsky_2019},
\begin{equation}
    x f_1^{\gamma/e}(x, k_\perp^2) = \frac{\alpha_{\text{eff}}}{\pi^2} \frac{k_\perp^2}{(k_\perp^2 + x^2 m_e^2)^2} \label{eq:photon-electron}
\end{equation}
where $ m_e $ is the electron mass, which is neglected. As shown in Fig.~1, the photon vertex on the initial electron involves the effective coupling $ \alpha_{\text{eff}} $. The distribution $ n(k) $ in $ \alpha_{\text{eff}} $ can also be obtained via the EPA.
\begin{equation}
    n(k_z, k_\perp^2) = 2(2\pi)^3 \frac{Z^2 \alpha_e}{\pi^2} \frac{k_\perp^2}{(k_\perp^2 + k_z^2/\gamma^2)^2} 
    \left[ F(k_\perp^2 + k_z^2/\gamma^2) \right]^2 \frac{k_z}{E_k}\label{eq:epa}
\end{equation}
where $\gamma = E_{\mathrm{nucleus}} / M_{\mathrm{nucleus}}\approx E_{proton} / M_{proton}$ is the Lorentz factor of the emitting particle. This factor $\gamma$ reflects the Lorentz contraction effect of the electromagnetic field at high velocities. Since the EPA flux is defined per unit $k_z$ and $k_\perp$, to match the Lorentz-invariant phase space element $\frac{d^3 k}{(2\pi)^3 2k_0}$, a factor of $2(2\pi)^3$ is introduced to yield the physical particle number distribution. $k_z/E_k$ is the Jacobian factor, which comes from the transformation.
\begin{equation}
\begin{aligned}
n(k_z, k_\perp^2) &= n(E_k, k_\perp^2) \frac{dE_k}{dk_z}= n(E_k, k_\perp^2) \frac{k_z}{E_k}
\end{aligned}
\end{equation}

\begin{figure}[htbp]
    \centering
    \begin{subfigure}{0.5\textwidth}
        \centering
        \includegraphics[width=\textwidth]{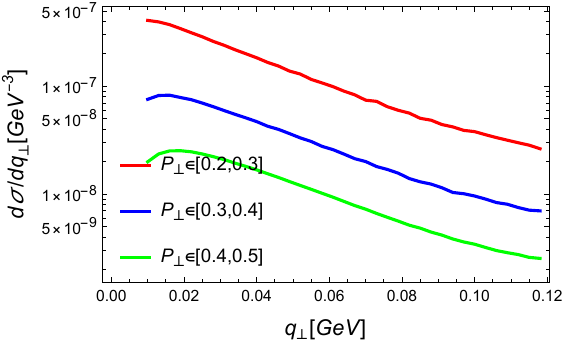} 
    \end{subfigure}\hfill
    \begin{subfigure}{0.49\textwidth}
        \centering
        \includegraphics[width=\textwidth]{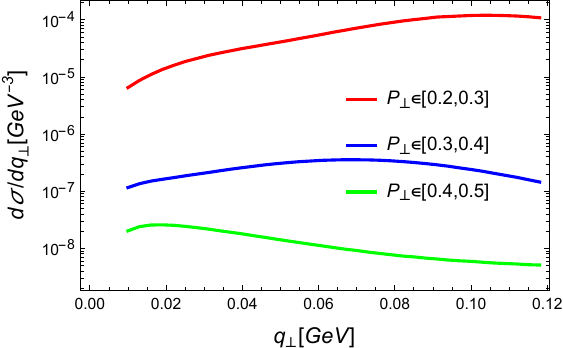} 
    \end{subfigure}
    \caption{Differential cross section for the process $e^-+Pb \to e^-+Pb+e^+e^-$ as a function of $ q_\perp $ at EicC energies. Left: without the background field effect. Right: with the background field effect. The rapidity of final-state electrons is integrated over the region [1.5, 2].}
    \label{f.fig2}
\end{figure}

\begin{figure}[htbp]
    \centering
    \begin{subfigure}{0.5\textwidth}
        \centering
        \includegraphics[width=\textwidth]{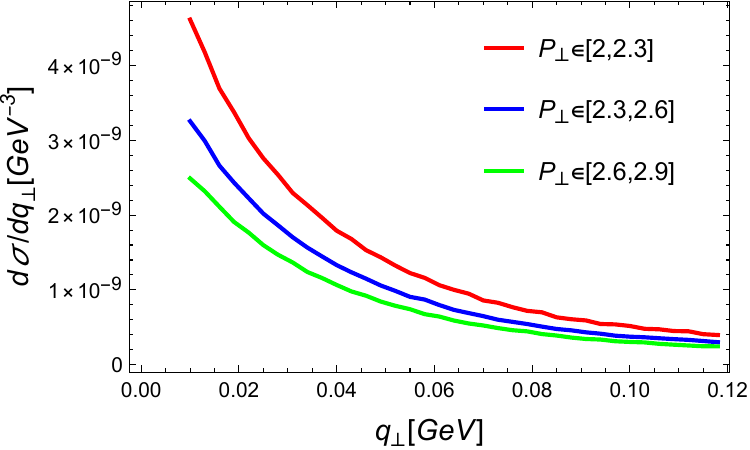} 
    \end{subfigure}\hfill
    \begin{subfigure}{0.49\textwidth}
        \centering
        \includegraphics[width=\textwidth]{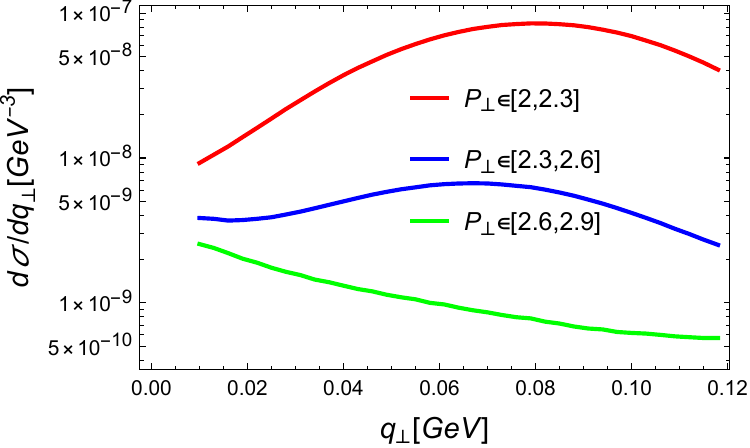} 
    \end{subfigure}
    \caption{Differential cross section for the process $e^-+Pb \to e^-+Pb+e^+e^-$ as a function of $ q_\perp $ at EIC energies. Left: without the background field effect. Right: with the background field effect. The rapidity of final-state electrons is integrated over the region [2, 2.8].}
    \label{f.fig3}
\end{figure}

To quantify the impact of background electromagnetic fields on quasi-real photon emission, we numerically evaluate the differential cross section $\frac{d\sigma}{dq_\perp}$ for lepton-pair production via two-photon fusion, both with and without the inclusion of background-induced corrections. 
For the EicC setup (with 3.5~GeV electron and 8~GeV heavy-ion beams), the final-state electron rapidity is integrated over the range $y \in [1.5,\, 2.0]$, with three representative transverse momentum intervals: $P_\perp \in [0.2,\, 0.3]$, $[0.3,\, 0.4]$, and $[0.4,\, 0.5]$~GeV. For the EIC setup (with 18~GeV electron and 100~GeV heavy-ion beams), the rapidity range is $y \in [2.0,\, 2.8]$, and the corresponding transverse momentum intervals are $P_\perp \in [2.0,\, 2.3]$, $[2.3,\, 2.6]$, and $[2.6,\, 2.9]$~GeV.
These kinematic regions are selected to ensure that the photon density is not so large as to make the effective coupling constant excessively strong, thus validating the use of perturbation theory. Figures~\ref{f.fig2} and~\ref{f.fig3} present the results for the EicC and EIC collider setups, respectively.

Numerical evaluations performed for the EicC setup reveal a strong sensitivity of the differential cross section to the background-field--induced coupling correction. 
As shown in Fig.~\ref{f.fig2}, the replacement of the vacuum coupling constant $\alpha_{e}$ by the effective coupling $\alpha_{\mathrm{eff}} $ leads to an enhancement of up to two orders of magnitude around $q_{\perp} \approx 0.01~\mathrm{GeV}$ across all examined transverse-momentum intervals.
The enhancement becomes increasingly significant with growing $P_{\perp}$, indicating that semi-hard photon emissions are particularly sensitive to the stimulated interaction effects in the dense nuclear electromagnetic field. 
In the large-$q_{\perp}$ region, the cross section exhibits an even more pronounced rise, which originates from the increased density of quasi-real photons in this kinematic domain. 
The higher photon occupation number effectively amplifies $\alpha_{\mathrm{eff}}$, thereby strengthening the overall interaction rate. 
These features collectively demonstrate that even within a purely QED process, strong background fields can substantially modify the photon phase-space distribution and reshape the momentum dependence of final-state lepton pairs.

At EIC energies (see Fig.~\ref{f.fig3}), a qualitatively similar but quantitatively milder trend is obtained. The background field effect leads to an enhancement of up to one order of magnitude around $q_{\perp} \approx 0.01~\mathrm{GeV}$ across all examined transverse-momentum intervals. 
The inclusion of the background correction produces a distinct flattening in the $q_{\perp}$ dependence between $0.02~\mathrm{GeV}$ and $0.08~\mathrm{GeV}$. 
This region coincides with the kinematic window most relevant to small-$x$ gluon measurements, implying that neglecting background-field effects could introduce systematic biases in the extraction of unintegrated gluon distributions from photon-initiated processes. 
Moreover, the modified $q_{\perp}$ spectrum, together with the observed redistribution of phase-space weight toward higher-momentum tails, indicates that the background field not only enhances the emission rate in the large-$q_{\perp}$ region but also alters the underlying transverse-momentum correlations, a feature that may be experimentally testable in precision lepton-pair correlation studies.

Additionally, we observe a qualitative modification in the $q_{\perp}$ dependence of the differential cross section. 
Without background correction, the cross section decreases mildly with increasing $q_{\perp}$; 
with correction, however, it becomes flatter or even slightly rising in some regions. 
This behavior indicates that the background field redistributes the available phase space at different transverse momenta, enhancing radiation in semi-hard and hard regions while depleting it at small $q_{\perp}$. 
Such nontrivial reshaping of the photon phase space may manifest in more differential observables, such as angular correlations or polarization asymmetries, providing a potential avenue for experimental verification.

In summary, our numerical results demonstrate that the background field effect induces non-negligible modifications to the cross section, especially in the semi-hard kinematic regions accessible at EIC and EicC. These effects should be systematically included in any precision program aimed at extracting the photon or gluon distribution functions from processes involving photon emission in background field. Moreover, the lepton-pair production process serves as a valuable control channel for testing initial-state photon flux models in realistic collider environments.

\section{Summary and outlook}
\label{3}

We have investigated the impact of background-field-induced modifications to the electromagnetic coupling on the extraction of small-$x$ gluon distribution functions in electron–ion collisions. Building on previous work in which the effective QED coupling $\alpha_{\text{eff}}$ was derived in the presence of classical electromagnetic fields, this study focuses on quantifying how such modifications propagate into observables sensitive to the initial-state photon flux.

To isolate and assess this effect, we analyzed the lepton-pair production process via quasi-real $\gamma\gamma$ fusion, which shares the same photon emission kinematics as the photon–gluon fusion channel $\gamma + g \rightarrow q\bar{q}$. Numerical calculations performed at EIC and EicC energies show that the inclusion of $\alpha_{\text{eff}}$ leads to a non-negligible enhancement in the differential cross section, particularly in the semi-hard transverse momentum region relevant for small-$x$ physics. These deviations are indicative of potential systematic shifts in gluon distribution extractions if background-induced QED corrections are neglected. Our results highlight the need to incorporate such effects in future high-precision analyses involving initial-state photon radiation at next-generation electron–ion colliders.

We emphasize that the background-field correction proposed in this work requires two essential conditions:(i) the process must involve a photon propagator, and (ii) the impact parameter must be sufficiently small so that the radiation takes place within a strong external electromagnetic field. In ultraperipheral $pp$ and $A+A$ collisions at the LHC, these conditions are not simultaneously fulfilled—photon emission occurs at large impact parameters and effectively in vacuum—so the background-field effect is expected to be negligible.

While this study establishes a theoretical framework and provides numerical evaluations for the background field effects on quasi-real photon emission, certain assumptions were adopted to simplify the problem. Specifically, we restricted the electron-radiated virtual photons to the quasi-real regime ($q^2\approx0$), which facilitates the effective use of the EPA for describing the background photon field. Although this approach highlights the core effect of background field corrections, future work should explore broader virtual photon momentum regions. Furthermore, the current model does not yet incorporate the background field's influence on the electron propagator, and the comprehensive inclusion of these effects will be a crucial direction for future investigations. Addressing these limitations, along with further analyses of more differential observables such as angular correlations and polarization asymmetries, will contribute to a more comprehensive understanding of background field effects and provide more refined theoretical support for future high-precision measurements at next-generation electron-ion colliders.





\section*{Declaration of generative AI and AI-assisted technologies in the writing process}

During the preparation of this work, the author(s) used ChatGPT to improve the linguistic quality, readability, and idiomatic expression of the manuscript. After applying this language assistance, the author(s) carefully reviewed and edited the content as needed and take(s) full responsibility for the final version of the publication.

\bibliographystyle{elsarticle-num} 
\bibliography{ref}






\end{document}